\documentclass{iopart}

\usepackage{graphicx}
\usepackage{bm}
\usepackage[latin1]{inputenc}
\usepackage{color}
\usepackage{amssymb}
\usepackage{times}
\usepackage{colordvi}
\usepackage{color}
\usepackage{epsfig}

\begin{document}

\title{The critical Kerr non-linear optical cavity in the presence of internal loss and driving noise}

\author{Andr\'e Th\"uring and Roman Schnabel}
\address{Max-Planck-Institut f\"ur Gravitationsphysik (Albert-Einstein-Institut) and\\
\mbox{Institut f\"ur Gravitationsphysik, Leibniz Universit\"at Hannover}\\
Callinstr.~38, 30167 Hannover, Germany}
\ead{Andre.Thuering@aei.mpg.de}

\date{August 06, 2011}

\begin{abstract}
We theoretically analyze the noise transformation of a high power continuous-wave light field that is reflected off a critical Kerr non-linear cavity (KNLC). Our investigations are based on a rigorous treatment in the time-domain. Thereby, realistic conditions of a {specific} experimental environment {including} optical intra-cavity loss and strong classical driving noise can be modeled for any KNLC. We show that even in the presence of optical loss and driving noise considerable squeezing levels can be achieved. We find that the achievable squeezing levels are not limited by the driving noise but solely by the amount of optical loss.  Amplitude quadrature squeezing of the reflected mean field is obtained if the KNLC's operating point is chosen properly. Consistently, a KNLC can provide a passive, purely optical reduction of laser power noise as experimentally demonstrated in~\cite{KhalaidovskiPRA80}. We apply our model to this experiment and find good agreement with measured noise spectra.   
\end{abstract}
\pacs{42.65.Hw, 42.50.Lc, 42.79.Gn}
\maketitle

\section{Introduction}
\begin{figure}[h!]
\begin{center}
\includegraphics[scale=1]{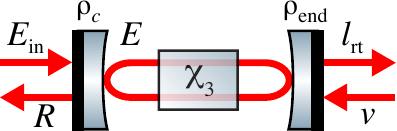}
\caption{ Schematic of a Kerr non-linear cavity (KNLC). $E_{\rm in}$ is the driving field coupled to the cavity, $E$ the intra-cavity field, $l_{\rm rt}$ the fraction of $E$ that is lost per round-trip at the imperfect end mirror and $v$ the vacuum field that couples in at the end-mirror. The reflected field is denoted as $R$. We investigated the regime where $\rho_{\rm c} <\rho_{\rm end}\leq 1$.}
\label{fig:KNLC}
\end{center}
\end{figure}

Due to the effect of self-squeezing the (quantum-) noise distribution of a light field can be manipulated when passed through a Kerr medium~\cite{KitagawaPRA34} or when coupled to and  reflected off a KNLC (Fig.~\ref{fig:KNLC})~\cite{DWa80kerr,CollettPRA32,PacePRA47,RehbeinPRL95}. In this way,  the optical Kerr effect can be used for the generation of squeezed light which can be exploited to enhance state-of-the-art metrology to quantum metrology. For a review we refer to~\cite{SchnabelNat}. In contrast to methods based on second order non-linear processes -- such as the optical-parametric-oscillation (OPO) -- no second harmonic pump field is required for operating the KNLC. Thus  a technical realization of squeezed light generation  seems to be less involved.  It was shown by Collet and Walls~\cite{CollettPRA32} that a loss-less KNLC operated at its so-called critical point (refer to the centered curves in  Fig.~\ref{fig:Intra}) provides perfect squeezing in the amplitude quadrature of the reflected field at zero Fourier frequency. This specific characteristic enables a further technical application in the classical noise regime:  In contrast to conventional techniques ~\cite{SeifertOL31, KweeOL33}, a KNLC can provide a light power stabilization within its bandwidth and in a purely optical, passive way, as it was observed in \cite{KhalaidovskiPRA80}. Such a passive, classical noise reduction scheme might find application in gravitational wave astronomy. Here, ultra-stable high-power laser radiation is needed~\cite{WillkeLPR} to achieve a sufficiently high interferometric sensitivity, e.g.~as envisioned for Advanced LIGO~\cite{aLIGO} and the Einstein Telescope~\cite{ET}.    

Reynaud \textit{et al.}~\cite{ReynaudPRA40} have analyzed the squeezing of quantum-noise achievable in reflection of a \emph{multi-stable} KNLC. They showed that a KNLC operated close to the turning points of the corresponding resonance curve also yields perfect squeezing at zero Fourier frequency, though not in the amplitude quadrature of the reflected field. In fact, they revealed that the amplitude quadrature at zero frequency is not squeezed if a loss-less multi-stable KNLC is considered. Consistently, the multi-stable state is not favorable if a passive, purely optical reduction of laser power noise is targeted at low frequencies.

In the pulsed laser regime the squeezing of quantum noise by means of the optical Kerr-effect was demonstrated in several experiments~\cite{BergmanOL16, DongOL33}. In these experiments no optical cavities were involved. The reduced quantum noise  was observed in certain field quadratures that depart from the amplitude quadrature of the mean field. Also in the continuous-wave laser regime one experiment was conducted to demonstrate squeezing of quantum noise \cite{ZhangPRA64}. Here a small reduction of about 1.5\,dB below the shot noise limit could be demonstrated by means of a KNLC driven by a 1.2\,mW laser beam. White \textit{et al.}~\cite{WhiteJOPTB2} have used a KNLC pumped by a 30\,mW laser beam and achieved a reduction by 1.5\,dB of \emph{classical noise}. Recently, at an intermediate-power 0.75\,W laser a classical power noise reduction by a surprisingly great factor of 32\,dB  by means of a KNLC was  demonstrated \cite{KhalaidovskiPRA80}. This strong reduction was observed at the driving laser's relaxation oscillation frequency where its noise in the amplitude quadrature is much higher than in the phase quadrature.

In this paper we report on our numerical investigations of the noise transformation of a high-power light field that is coupled to and reflected off a critical KNLC. 
{We used the approach of a rigorous treatment in the time-domain which easily includes the full nonlinear formalism and the full range of different experimental parameters such as low and high cavity finesse values.}  It is thus also possible to calculate the cavity dynamics on time-scales of the cavity storage time revealing cavity ringing or the hysteresis effect, as well as the non-linear response to large (classical) signals, refer e.g.~to Fig.~6.15 and Fig.~6.17 in \cite{Thuering09}, respectively. {Here}, we use our model to describe the squeezing of quantum noise and {weak} classical {driving noise with Gaussian statistics under consideration of the KNLC's detuning and loss.}  Model input parameters are the cavity length, cavity mirror reflectivities, the intra-cavity optical loss, the Kerr non-linearity and the optical pump power.   As all of these parameters can be determined or at least estimated in experimental environments, our numerical calculations have a great potential to easily model and understand experimental results, e.g.~those reported in ~\cite{KhalaidovskiPRA80}.

In the following, the noise distribution in phase-space is analyzed for a quantum noise limited driving field  as well as for a field that shows {(unbalanced and possibly correlated)} classical noise in its amplitude and phase quadrature. In both cases, several values for the internal cavity loss are considered, and for certain operating points the noise transformation is illustrated  by the corresponding Wigner functions~\cite{WignerPhysRev40}.
We focus on the noise reduction in the amplitude quadrature of the mean field  that is reflected off the KNLC. This is essential in view of a potential passive, purely optical reduction of laser power noise.  We show that the presence of intra-cavity loss strongly influences the phase-space rotation and thus the quadrature yielding the optimal noise reduction.  Additionally, we show that even for a driving field with an {unbalanced} (classical) noise distribution in its amplitude and phase quadratures strong noise reductions, even below quantum noise, can be achieved. As already shown by Collet and Walls~\cite{CollettPRA32} for a quantum noise limited driving field, the loss-less KNLC needs to be operated at its critical point in order to obtain optimal squeezing (noise reduction) in the amplitude quadrature of the reflected mean field. We show that this condition still holds for a driving field that exhibits classical noise in both amplitude and phase quadrature. In the presence of intra-cavity loss, however, amplitude quadrature noise reduction can be obtained with a critical KNLC only if operated with a detuning aside from the critical point.

\section{Calculation of the light fields}

The non-linear nature of a KNLC becomes evident by looking at the {analytic expression for the monochromatic, steady-state intra-cavity field (for zero loss), as} given by
\begin{equation}
E = \frac{\mathrm i\tau_{\rm c}}{1-\rho_{\rm c}\rho_{\rm end}\exp[2\mathrm i(\Phi+\theta |E|^2)]}E_{\rm{in}}\,. \label{eq:Intra}
\end{equation}
Here $\rho_{\rm c}$ ($\tau_{\rm c}$) and $\rho_{\rm end}$ ($\tau_{\rm end}$) are the amplitude reflectance (transmittance) factors of the coupling and end mirror, respectively. $\Phi$ denotes the {geometric} detuning with respect to the carrier light frequency $\omega_{0}$. Furthermore, $\theta|E|^2$ is the intensity dependent phase shift induced by the optical Kerr effect. In the variable $\theta$  the non-linear refraction index $n_{2}$, the length of the Kerr medium $L_{\rm{KM}}$, the cross sectional area $\mathcal{A}$ of the light field, the speed of light $c$ and the carrier frequency $\omega_{0}$ are included according to \cite{RehbeinPRL95}
\begin{equation}
\theta = \frac{n_{2}\omega_{0}L_{\rm{KM}}}{2\mathcal{A}c}\,.\label{eq:theta}
\end{equation}

\begin{figure}
\begin{center}
\includegraphics[scale=.75]{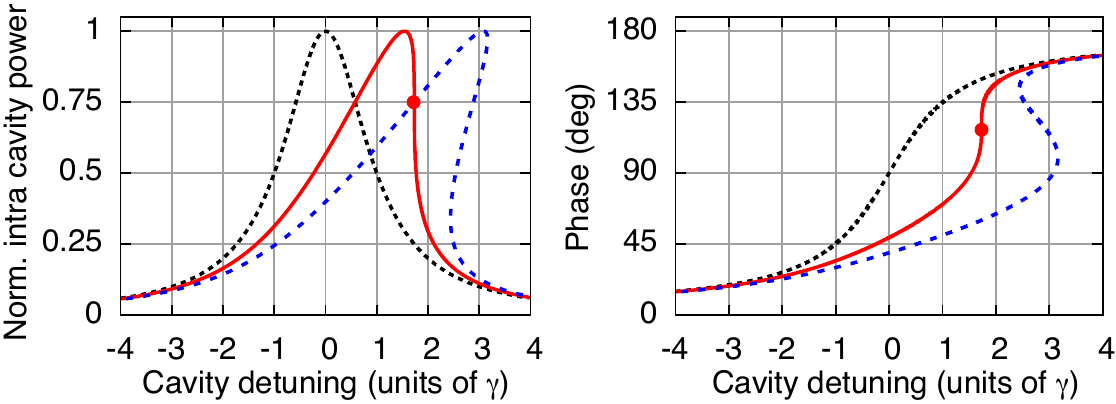}
\caption{ Normalised intra cavity power (left) and phase (right) without Kerr effect (black dotted curves), the so-called critical state (red solid curves) with its critical point of infinite slope (highlighted by the red circles)  and a multi-stable state (blue dashed curves) of the KNLC with half-bandwidth $\gamma$.}
\label{fig:Intra}
\end{center}
\end{figure}

Fig.~\ref{fig:Intra} shows the intra-cavity power $P= |E|^2$ depending on the (static) cavity detuning $\Phi$ for three cases: the linear case with $\theta = 0$ (black dotted curve), the so-called critical state with  $\theta = \theta_{\rm{crit}}$ (red solid curve) and a multi-stable case with ${\theta =2 \theta_{\rm{crit}}}$ (blue dashed curve).   As Eq.~(\ref{eq:Intra}) is a transcendental (implicit) equation, it {has been solved numerically}. Fig.~\ref{fig:Intra} reveals that the critical state and the multi-stable state exhibit particular operating points  at which the resonance curve shows an infinite slope. In the critical state the infinite slope occurs at the critical operating point ${\rm{OP}}_{\rm{crit}}$ (highlighted by the red circles) and in the multi-stable state at the two turning points of the resonance curve. As shown in \cite{CollettPRA32} and \cite{ReynaudPRA40} these operating points are favorable in view of the {achieving high} squeezing levels. However, since in the multi-stable case perfect squeezing is not obtained in the amplitude quadrature of the reflected field we restrict our investigations to the critical state. 

In order to {model noisy optical fields reflected off the KNLC}  
in the time-domain it is necessary to evaluate the light fields (and their interference) after each round-trip. As we are interested in the noise transfer of a KNLC set to a certain operating point, we assume a static detuning of the cavity, i.e. $\Phi(t)=\Phi={\rm{constant}}$. Following this line, the internal field reads
\begin{equation}
\label{eq:intra}
 E(n t_{\rm{rt}})=\mathrm i\tau_{\rm c}E_{\rm{in}}(n t_{\rm{rt}}) + E_{\rm rt}\left[(n-1) t_{\rm{rt}}\right]\,.
 \end{equation}
Here 
\begin{eqnarray}
\nonumber& & E_{\rm rt}\left[(n-1) t_{\rm{rt}}\right]=\\
& &\rho_{\rm end}\rho_{\rm c}\exp\left[2\mathrm i\Phi+2\mathrm i\theta\left|E\left[\left(n-1\right)t_{\rm{rt}}\right]\right|^2\right] E\left[(n-1)t_{\rm{rt}}\right]\label{eq:t2mc_dynamic}
\end{eqnarray}
describes the field after one round-trip. 
{The nonlinearity of our approach at this stage is evident due to the exponential function in Eq.~(\ref{eq:t2mc_dynamic}) depending on the field intensity $|E|^2$.}
The round-trip time in a cavity of length $L$ is
\begin{equation}
t_{\rm{rt}} = \frac{2L}{c}\,.
\end{equation}
The reflected field is then given by
\begin{equation}
\label{eq:refl}
R(n t_{\rm{rt}}) = \rho_{\rm c}E_{\rm{in}}(n t_{\rm{rt}}) +{\rm i} \tau_{\rm c}E(n t_{\rm{rt}})\,.
\end{equation}
The time dependence of the input field can be described by 
\begin{equation}
\label{eq:ein}
E_{\rm{in}}(t) = \overline{E_{\rm{in}}}+ \delta E(t)_{\rm{in}}\quad\rm{with}\quad\overline{\delta E_{\rm{in}}(t)} = 0\,.
\end{equation}
If the mean field $\overline{E_{\rm{in}}}$ is {set} to be real,  Eq.~(\ref{eq:ein}) can be written as
\begin{equation}
\label{eq:input}
E_{\rm{in}}(t) = E_{0}+\delta X_{1}(t) +{\rm i}\delta X_{2}(t)
\end{equation}
giving a real-valued phase-space description of the {input} laser field with $\delta X_{1}(t)$ and $\delta X_{2}(t)
$ being the fluctuations in its amplitude ($X_1$) and phase quadrature ($X_2$), respectively. 
{The same description does then apply to the reflected field after $n$ round-trips according to Eq.~(\ref{eq:refl}) and its converged value for $n \Rightarrow \infty$.}

{Eventually, the cavity reflected fluctuating fields of our converged time-domain simulations need to be expressed as normalized noise spectral densities in order to allow for a comparison with experimental data. For this reason we apply to the reflected field a fast Fourier transformation (FFT) and normalize the result such that, for a zero Kerr effect, the vacuum noise of a coherent input field transforms into a noise spectral density of magnitude unity (0\,dB) with a white spectrum. }

{In the following we describe our time-domain simulation in more detail. We also describe the transfer functions we used to calculate the noise spectral densities for other than vacuum noise input, i.e.~for classical noise inputs with unbalanced, possibly correlated, fluctuations in the amplitude and phase quadratures. With respect to the transfer functions we restrict ourselves to the regime where the noise amplitude is much smaller than the driving field amplitude, i.e.~to the regime where a linearized approximation is valid. For all results presented here, a fluctuation at a certain Fourier frequency is always transformed into output noise at the same single frequency, since the Fourier transformation does not show any additional frequency components. Note that we indeed have observed additional frequency components for parameters outside the regime presented in this manuscript.} 

We consider an amplitude-modulated input field by setting Eq.~(\ref{eq:input}) to 
\begin{equation} 
E_{\rm in} = E_{\rm am}(t) = E_{0}+x_{1} \cos(\Omega t)
\end{equation}
and accordingly to
\begin{equation} 
E_{\rm in} = E_{\rm pm}(t) = E_{0}+{\rm i}x_{2} \cos(\Omega t)\,.
\end{equation}
for a phase-modulated field. 
{Here, $\Omega$ is the angular Fourier frequency and $x_1$ and $x_2$ are real-valued scaling factors with $|x_1|, |x_2| \ll |E_0|$.} 
{The FFT (we have used FFTW3~\cite{FFTW05}) of the reflected field $R(nt_{\rm rt})$ provides the relative amplitudes of upper and lower sidebands at positive and negative frequencies  $R(\pm\Omega)$, respectively.  The phases of upper and lower sidebands determine whether an input amplitude modulation is transferred to an output amplitude or phase modulation. The full coupling can be described by 4 coefficients $T_{ij}$
}
\begin{eqnarray}
T_{11} &=&\left[R_{\rm am}^*(-\Omega)+R_{\rm am}(\Omega)\right]/(2x_1) \label{eq:t11}\\
T_{12} &=& \left[R_{\rm pm}^*(-\Omega)+R_{\rm pm}(\Omega)\right]/(2x_2)\\
T_{21} &=& \left[{\rm i}[R_{\rm am}^*(-\Omega)-R_{\rm am}(\Omega)]\right]/(2x_1)\\
T_{22} &=& \left[{\rm i}[R_{\rm pm}^*(-\Omega)-R_{\rm pm}(\Omega)]\right]/(2x_2)\,.\label{eq:t22}
\end{eqnarray} 
These coefficients can be written as a $2\times2$-matrix {$\mathbf{T}(\Omega)$, which is commonly referred to as the (linearized) input-output transfer function}. 
The spectral density matrix (covariance matrix) reads
\begin{equation}
\label{eq:spectral}
\mathbf{S}(\mathbf{T}) = \frac{1}{2}\left(\mathbf{T}\cdot\mathbf{T}^\dagger+\mathbf{T}^*\cdot\mathbf{T}^{\rm T}\right)\,.
\end{equation}
The diagonal components of the $2\times2$-matrix $\mathbf{S}$ correspond to the {normalized  power} spectral densities of the field's amplitude ($X_1$) and phase ($X_2$) quadrature amplitudes. Off-diagonal components are due to correlations between the two quadratures. From this matrix the spectral density of any measured linear combination \mbox{$X_\zeta = \cos(\zeta)X_1+\sin(\zeta)X_2$} can be evaluated to
\begin{equation}
S_\zeta = \begin{pmatrix}{\cos\zeta \;\;  \sin\zeta}\end{pmatrix}\cdot \mathbf{S}(\mathbf{T})\cdot\begin{pmatrix}{\cos\zeta ; \; \sin\zeta}\end{pmatrix}\,,\label{eq:quadnoise}
\end{equation}
where $\zeta$ denotes the homodyning angle. 
If the noise transformation is considered  for other than {the reference} vacuum input, the matrix $\mathbf{T}$ needs to be replaced by
\begin{equation}
\label{eq:classnoise}
\mathbf{T^\prime} = \mathbf{T}\cdot\begin{pmatrix}{\cos\vartheta \; -\sin\vartheta ; \;\sin\vartheta \;\;  \cos\vartheta}\end{pmatrix}\cdot\begin{pmatrix}{\exp(2r_1) \;\;  0 ;\;
0  \;\; \exp(2r_2)}\end{pmatrix}\,.
\end{equation}
The matrix on the right is the general squeezing matrix {with $(r_1+r_2) \geq 0$ due to Heisenberg's uncertainty relation}. Values $r_{1,2} < 0$ correspond to a noise level below vacuum noise (squeezed) and $r_{1,2} > 0$ above vacuum noise (anti-squeezed), respectively. The matrix on the left describes a rotation of the squeezing (noise) ellipse in phase space, i.e. the squeezed quadrature is determined by the squeezing angle $\vartheta$. In order to constitute  a driving field that exhibits  stationary classical noise (such as thermal noise) in its amplitude and phase quadrature,   both $r_1$ and $r_2$ need to be greater than zero. 
{For an extensive overview of this linear transfer function formalism we refer to \cite{HarmsPHD}.}

{
In addition to quantum noise and classical driving noise that enters the KNLC through the coupling mirror, also vacuum noise contributions due to intra-cavity loss need to be considered and 
}
the corresponding input-output relation (denoted as  $\mathbf{L}$ in the following) derived. For simplicity, we map the cavity round-trip loss $l_{\rm rt}$ onto the end mirror amplitude transmissivity  $\tau_{\rm end}=l_{\rm rt}$.  Hence, similar to the derivation of the input-output relation $\mathbf{T}$ we consider the transfer function for the following light fields
\begin{eqnarray}
v_{\rm am}(t) &=& x_1\cos(\Omega t)\\
v_{\rm pm}(t) &=& {\rm i}x_2\cos(\Omega t)
\end{eqnarray}
that couple in at the end mirror. {The beat with the driving field is described by setting $E_{\rm in}(t)$ to $E_0$, and the time-dependent loss-driven amplitude modulation inside the cavity then reads}
\begin{eqnarray}
E_{\rm am}(nt_{\rm rt}) & = &  {\rm i}\tau_{\rm end}v_{\rm am}(t) + {\rm i}\tau_{\rm c} E_0\exp\left[{\rm i}\Phi_{\rm opt}[(n-1)t_{\rm rt}]\right]\nonumber\\
& &   + E_{\rm rt, am}[(n-1)t_{\rm rt}]\,.
\end{eqnarray}
The contribution to the overall noise is given by the KNLC transmitted part, which reads
\begin{equation}
T(nt_{\rm rt}) = {\rm i}\tau_{\rm c}E_{\rm am}(nt_{\rm rt})\exp\left[{\Phi_{\rm opt}}[(n-1)t_{\rm rt}]\right]\,.
\end{equation}
In analogy to Eqs.~(\ref{eq:t11}) -- (\ref{eq:t22}), the coefficients of the input-output {transfer function}  $\mathbf{L}(\Omega)$ are determined from the FFT of $T(nt_{\rm rt})$. Finally, the total {power} spectral density matrix is given by
\begin{equation}
\mathbf{S}_{\rm tot}(\Omega)=\mathbf{S}(\mathbf{T}(\Omega))+ \mathbf{S}(\mathbf{L}(\Omega))\,.\label{eq:allnoise}
\end{equation}

\section{Amplitude quadrature squeezing}

\begin{figure}[tp]
\begin{centering}
\includegraphics[scale=0.71]{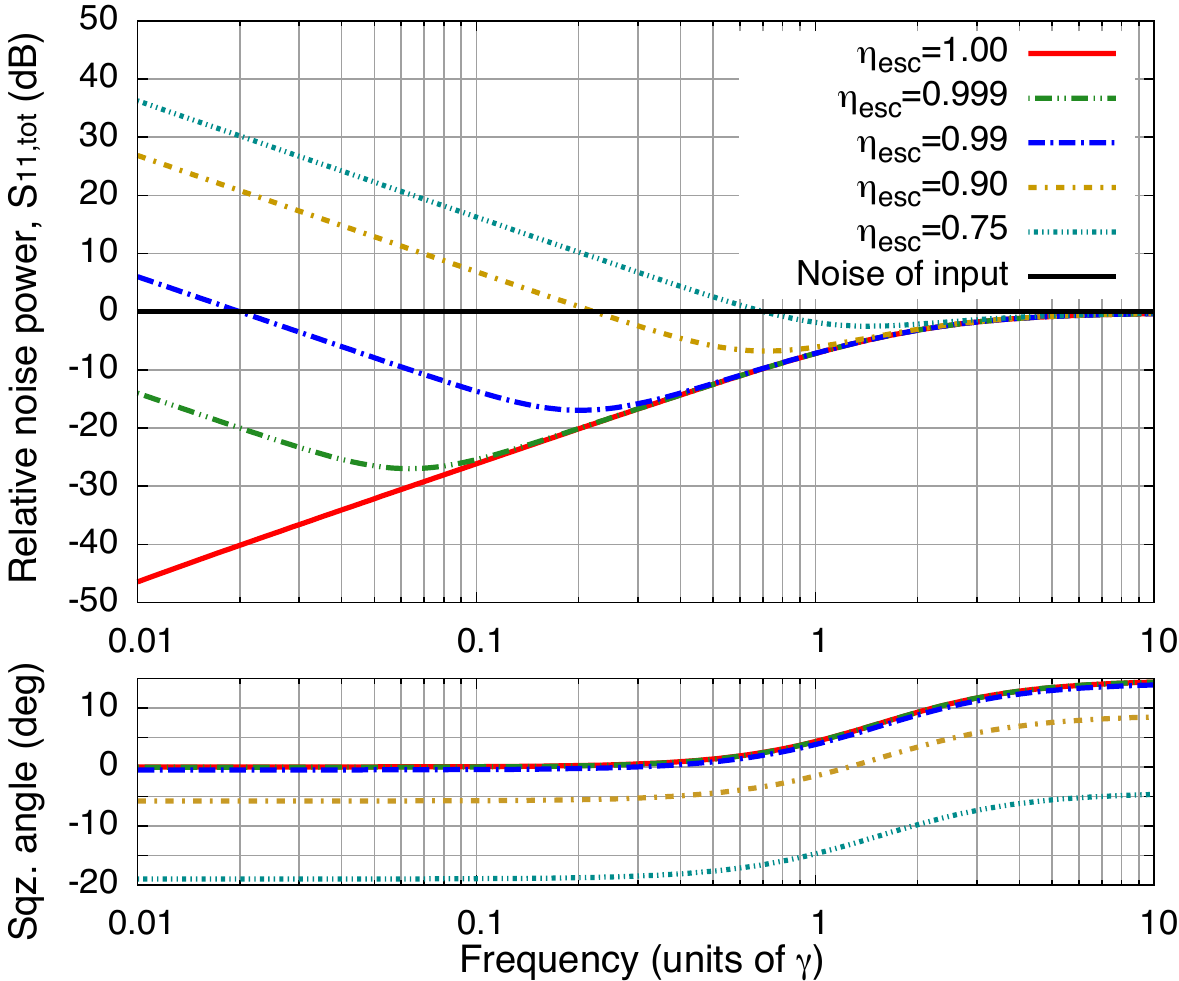}
\caption{ \textbf{Top:} Amplitude quadrature noise spectra of the light field reflected off the critical KNLC for a shot noise limited input field (0\,dB reference). The frequency is normalized to the respective cavity half-bandwidth $\gamma$. Five values for the escape efficiency $\eta_{\rm{esc}}$  are considered.  In all cases squeezing of quantum noise can be observed. \textbf{Bottom:} The graph shows the frequency dependent squeezing angle. Only for the loss-less case the squeezed quadrature matches the $X_1$-quadrature ($0^\circ$) at low frequencies. In presence of intra-cavity loss the squeezed quadrature deviates from the $X_1$-quadrature. This deviation increases with increasing loss.}
\label{fig:specx1}
\end{centering}
\end{figure}

First, we analyze a shot noise limited (vacuum noise limited, $r_1=r_2=0$) input field and calculate the {power} noise spectra for several values of the KNLC round-trip loss $l_{\rm rt}^2$. It is convenient to quantify the loss in terms of the so-called escape efficiency
\begin{equation}
\eta_{\rm{esc}} = \frac{\tau_\textrm{c}^2}{l_{\rm rt}^2+\tau_\textrm{c}^2}\quad\Leftrightarrow\quad l_{\rm rt}^2=\frac{\tau_\textrm{c}^2(1-\eta_{\rm{esc}})}{\eta_{\rm{esc}}}\,.
\end{equation}
We consider escape efficiencies of 1  (loss-less), 0.999, 0.99, 0.9 and 0.75, respectively. In all cases, the KNLC is set to its critical point and the power noise spectrum is calculated for the $X_1$-quadrature of the reflected mean field {($S_{11,\rm tot}$)}. The results are shown in Fig.~\ref{fig:specx1}. The spectrum obtained for the loss-less case agrees with the results of Collet~\cite{CollettPRA32} and Reh\-bein~\cite{RehbeinPRL95} and  yields the predicted squeezing in the amplitude quadrature. If a KNLC with internal optical loss ($\eta_{\rm esc}<1$) is considered, the spectra show a qualitatively different behaviour. The intra-cavity loss affects the phase space rotation of the reflected light field such that the squeezed quadrature deviates from the amplitude quadrature. This deviation increases with increasing loss as can be seen in the lower graph of Fig.~\ref{fig:specx1}. Please refer also to the Wigner functions obtained for the critical point (OP$_{6,\rm crit}$) shown in  Figs.~\ref{fig:Wig01gamma} and \ref{fig:Wiggamma}. Correspondingly, for comparatively  low escape efficiencies (cyan long dash-dotted and yellow dash-dotted curves in Fig.~\ref{fig:specx1})  the beam reflected off the KNLC shows a considerably enhanced amplitude quadrature noise at low frequencies. Around the pole frequency of the KNLC  the  cavity dispersion turns the noise ellipse and thus the squeezed quadrature towards the amplitude quadrature of the reflected  light field. As the \mbox{(anti-)squeezing} levels degrades at frequencies above the cavity's bandwidth, only moderate squeezing levels can be achieved around $\Omega=\gamma$.  In the case of high escape efficiencies ($\eta_{\rm esc}=0.999$ and 0.99)  considerable amplitude quadrature squeezing can still be achieved in the mid frequency range.  On the contrary, at  low frequencies where  the  (anti-)squeezing factors are comparatively high, already a small deviation of the squeezed quadrature from the amplitude quadrature leads to a degraded squeezing level or even a noise enhancement, respectively.

\begin{figure}[h!!]
\begin{center}
\includegraphics[scale=0.71]{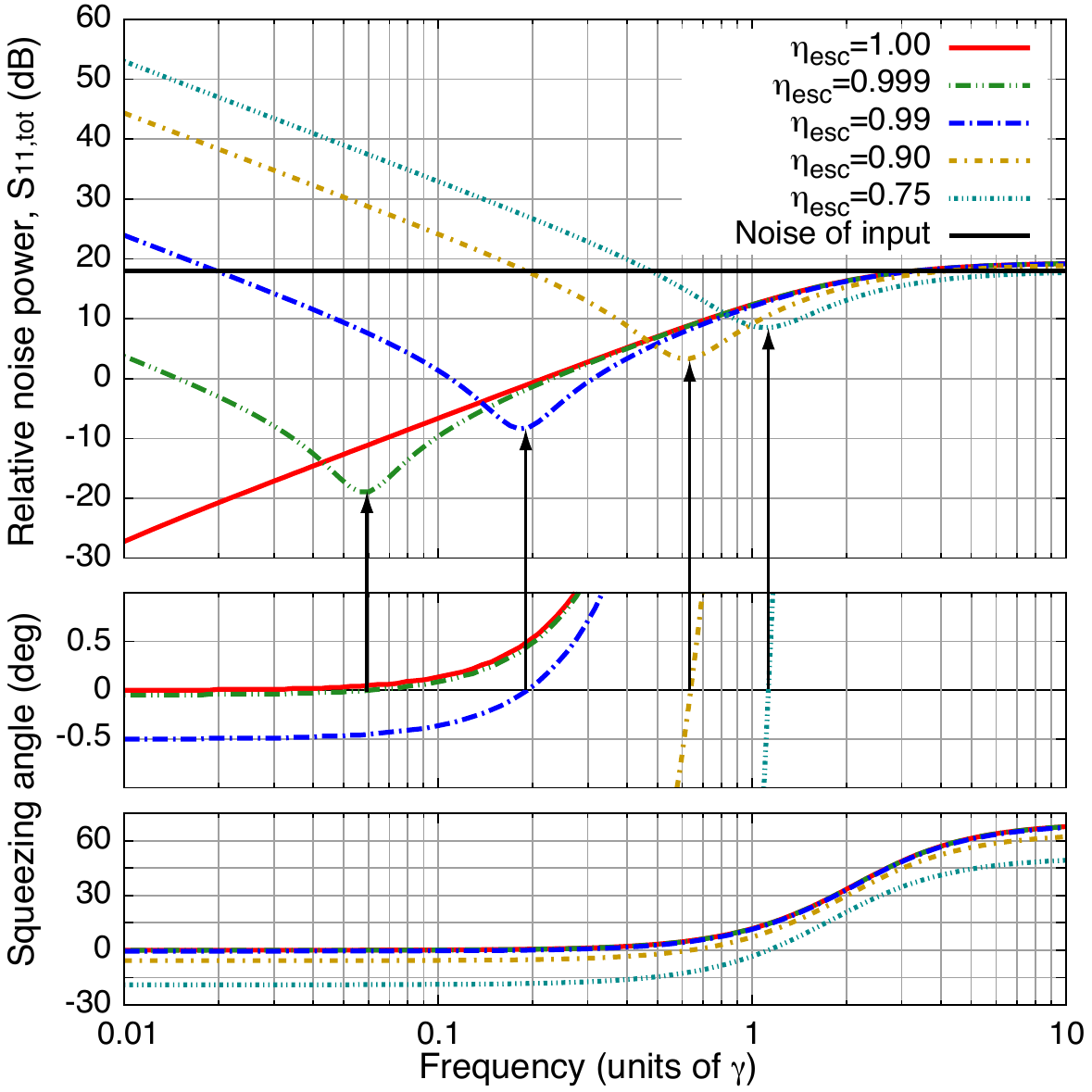}
\caption{ \textbf{Top:}  Amplitude quadrature noise spectra of the KNLC as shown in Fig.~\ref{fig:specx1} but for an input field with a classical and unbalanced noise distribution in quadrature phase-space. The noise power in the $X_{\vartheta}$-quadrature and in the $X_{\vartheta+90^\circ}$-quadrature are set to  20\,dB and 10\,dB above shot noise, respectively. The quadrature angle $\vartheta$ is set to $40^\circ$. \textbf{Middle, Bottom:} The frequency dependent quadrature angle yielding the lowest noise level. The middle graph shows a zoom of the bottom graph to make the origin of the dips in the spectra obtained for $\eta_{\rm esc}\neq 1$ more obvious (see text). At low frequencies this quadrature angle is almost the same as in Fig.~\ref{fig:specx1}. At frequencies  far above the half-bandwidth $\gamma$ the squeezing factors are comparatively small but the initial noise distribution is still rotated in phase-space due to the detuned KNLC.}
\label{fig:arbspecx1}
\end{center}
\end{figure}

Likewise, we analyze the noise transformation for an input field that {still has a Gaussian statistics but} exhibits an unbalanced noise distribution in phase-space. To constitute a realistic situation of an experiment we choose the noise of the driving field in its $X_{\vartheta}$-quadrature  with 20\,dB and in its $X_{\vartheta+90^\circ}$-quadrature with 10\,dB above {the reference value}. The quadrature angle  $\vartheta$ is set to $40^\circ$ with respect to the \mbox{$X_1$-quadrature} of the driving mean field. The results obtained are shown in Fig.~\ref{fig:arbspecx1}. In the loss-less case ($\eta_{\rm esc}=1$) the relative spectral noise reduction is essentially the same as in Fig.~\ref{fig:specx1}. At high frequencies a slightly  enhanced noise level appears. Here, the (anti-)squeezing levels are almost zero but the cavity dispersion causes a rotation of the initial noise ellipse (refer to the lower graph) such that the $X_\vartheta$-quadrature approximates the $X_1$-quadrature of the reflected mean field. Please note that at the lower bound of the considered spectrum the noise reduction is about 45\,dB just as in the case of a shot noise limited driving field (refer to Fig.~\ref{fig:specx1}). The squeezing is still about 30\,dB although the driving field exhibits significant classical noise in both of its quadratures.    
In contrast to Fig.~\ref{fig:specx1} the spectra obtained for a KNLC with ${\eta_{\rm esc} \neq 1}$ show narrow dips that fall below the noise level achieved in the loss-less case. This feature becomes evident by looking at the middle graph of Fig.~\ref{fig:arbspecx1}. It shows the frequency dependence of the quadrature that yields the lowest noise level. The dips in the spectra occur at that frequency where the squeezed quadrature coincides with the \mbox{$X_1$-quadrature} (i.e. with $0^\circ$) of the reflected mean field. This fact is highlighted by the arrows pointing from the middle towards the top graph.

\section{Influence of the cavity operating point}

\begin{figure}[t]
\begin{center}
\includegraphics[scale=0.245]{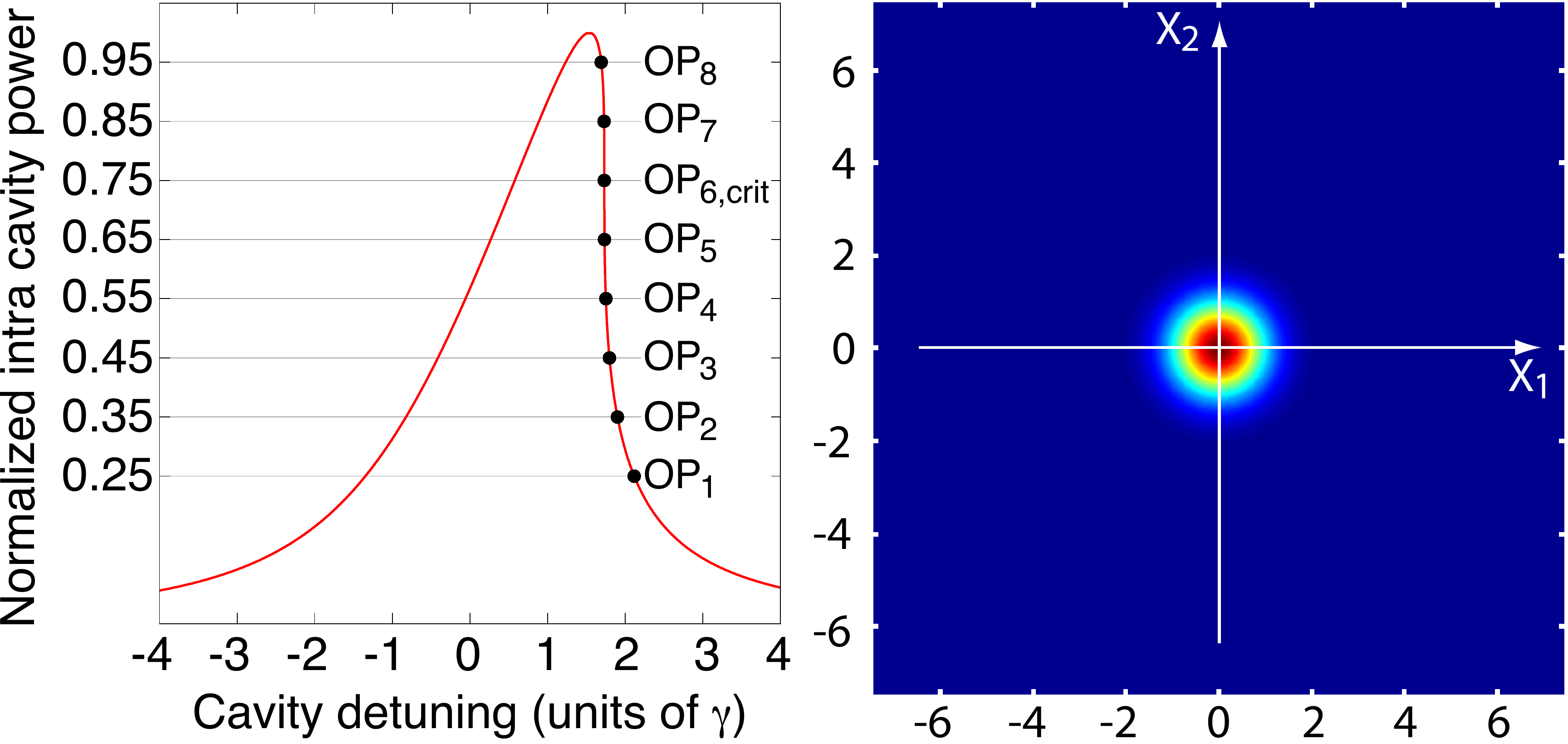}
\caption{ \textbf{Left:} The graph shows eight OPs on the steep resonance slope of a critical KNLC. The operating point OP$_{6, \rm{crit}}$ is the critical point. For all OPs the noise transformations are investigated in phase space at sideband frequencies corresponding to $0.1\gamma$ and $\gamma$, respectively. \textbf{Right:} The graph shows the Wigner function of a vacuum state providing a reference for our analysis.  Note that all modulation fields at sideband frequencies of a monochromatic coherent state carrier field are, per definition, in a vacuum state. The $X_1$-axis corresponds to the amplitude quadrature of the carrier field, and the $X_2$-axis to its phase quadrature, respectively.}
\label{fig:WigOps}
\end{center}
\end{figure}

The foregone investigations shows that for intra-cavity loss (\mbox{$\eta_{\rm esc}< 1$}) the squeezed quadrature does not coincide with the \mbox{$X_1$-quadrature} of the reflected mean field at low frequencies. It is obvious that this quadrature angle must have a connection with the chosen operating point (OP) of the KNLC. Since we want to prove the KNLC's capability for a passive, purely optical reduction laser power noise under consideration of a realistic experimental situation, we investigate the noise transformation in dependence of the OP. In view of a possible application in a high-power laser stabilization for e.g.~advanced gravitational-wave detectors~\cite{ET, aLIGO} a noise reduction at frequencies down to 1\,Hz are required. Thus, an optimization with regard to the noise reduction at low frequencies is favorable. We investigate the noise transformation for eight OPs at certain frequencies ($\Omega=0.1\gamma$ and $\Omega=\gamma$, where $\gamma$ is the KNLC's half-bandwidth).  The considered OPs lie on the steep resonance slope of the KNLC as illustrated in the left graph of Fig.~\ref{fig:WigOps}. They correspond to 0.25 to 0.95 of the intra-cavity power $P_{\rm res}$ that is achieved on resonance. In all cases  the critical point relates to OP$_{6, \rm crit}$ yielding  an intra-cavity power of $0.75P_{\rm res}$  \cite{Lastzka10}. 

{Generally, the modulation state of a light field at Fourier frequency $\Omega$ is fully characterized by its quasi-probability distribution in the quadrature-amplitude phase space -- the so-called Wigner function. Plotted for several OPs, the Wigner function nicely illustrates the different noise transformations possible for a KNLC. 
}
{Within the linearized approximation, Gaussian input noise is generally transformed into noise with again Gaussian statistics.} 
For such states the corresponding Wigner function is determined by the maximal and minimal noise levels (in the two orthogonal quadratures) and the orientation of the noise ellipse in phase-space. One obtains
\begin{equation}
W= \frac{1}{\pi} \exp[-x_{1,\vartheta}^2\exp(2r_1)-x_{2,\vartheta} ^2\exp(2r_2)]\label{eq:wig}
\end{equation}
with
\begin{eqnarray}
x_{1,\vartheta} &=& x_1\cos\vartheta - x_2\sin\vartheta\\
x_{2,\vartheta}  &= &x_1\sin\vartheta + x_2\cos\vartheta\,.
\end{eqnarray}
The factors $\exp(2r_{1,2})$ accounts for a squeezed (${r_{1,2}<0}$) or an anti-squeezed noise (${r_{1,2}>0}$). For $r_{1,2}=0$ one obtains the Wigner function of a  vacuum state as shown in the right graph of Fig.~\ref{fig:WigOps}. The Wigner function of a pure, 10\,dB amplitude-squeezed  state is determined by ${r_1=\ln(0.1)/2}$, ${r_2=\ln(10)/2}$ and ${\vartheta=0}$. 

We note that Kitagawe and Yamamoto~\cite{KitagawaPRA34} and Reynaud \textit{et al.}~\cite{ReynaudPRA40} showed that non-Gaussian Wigner functions can be obtained for \emph{strong} Kerr non-linearities at zero frequency. 
{
In the frequency and Kerr nonlinearity regime investigated here, non-linear transfer functions and thus non-Gaussian Wigner functions would also occur, but only for strongly mixed input states with a corresponding quadrature noise many orders of magnitude above shot noise. The result would be a \emph{mixed} non-Gaussian state with an entirely positive Wigner function. In this work, however, we apply our model to the regime where the noise amplitude is much smaller than the amplitude of the mean driving field. 
} 
For every calculation, we have verified that the transfer functions and thus the noise transformation are indeed linear, thus, the approximation of Gaussian Wigner functions is appropriate. 

\begin{figure*}
\begin{center}
\includegraphics[scale=0.31]{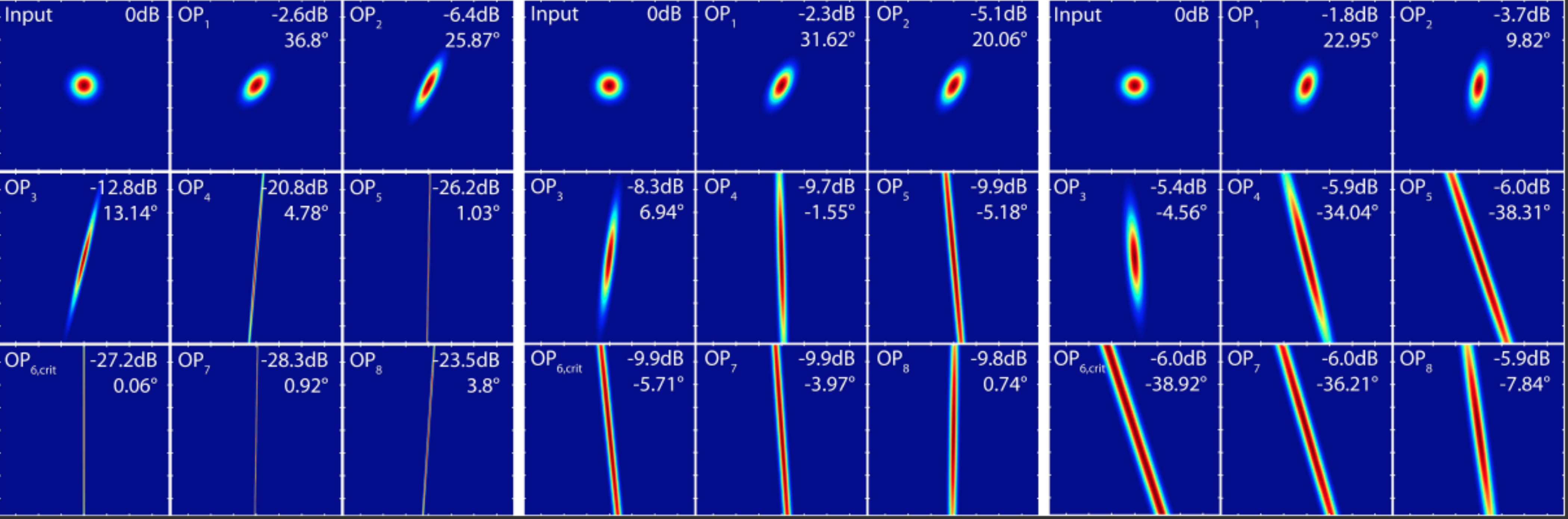}
\caption{ Noise transformations of a (pure) vacuum input state at sideband frequency {$\Omega=0.1\gamma$} for several operation points (OPs) of the KNLC, in terms of Wigner functions. Note that every Wigner function is normalized to its maximum. By that, the size of the colored area increases with the mixedness of the state. For every Wigner function the angle and the vacuum normalized variance of the state's lowest noise quadrature  is given. \textbf{Left:} $\eta_{\rm esc} = 1$ (loss-less). At OP$_{\rm6,crit}$ the squeezed quadrature matches the amplitude quadrature.  \textbf{Middle:} $\eta_{\rm esc} = 0.9$. The loss changes the phase-space rotation such that the squeezed quadrature matches the $X_1$-quadrature at \emph{two} OPs, one close to OP$_4$ and another close to OP$_8$.  \textbf{Right:} $\eta_{\rm esc} = 0.75$. Again, the squeezed quadrature matches the $X_1$-quadrature at OPs that depart from the critical point OP$_{6,\rm{crit}}$. The high optical loss leads to strongly mixed states.}
\label{fig:Wig01gamma}
\end{center}
\end{figure*}

\begin{figure*}
\begin{center}
\includegraphics[scale=0.31]{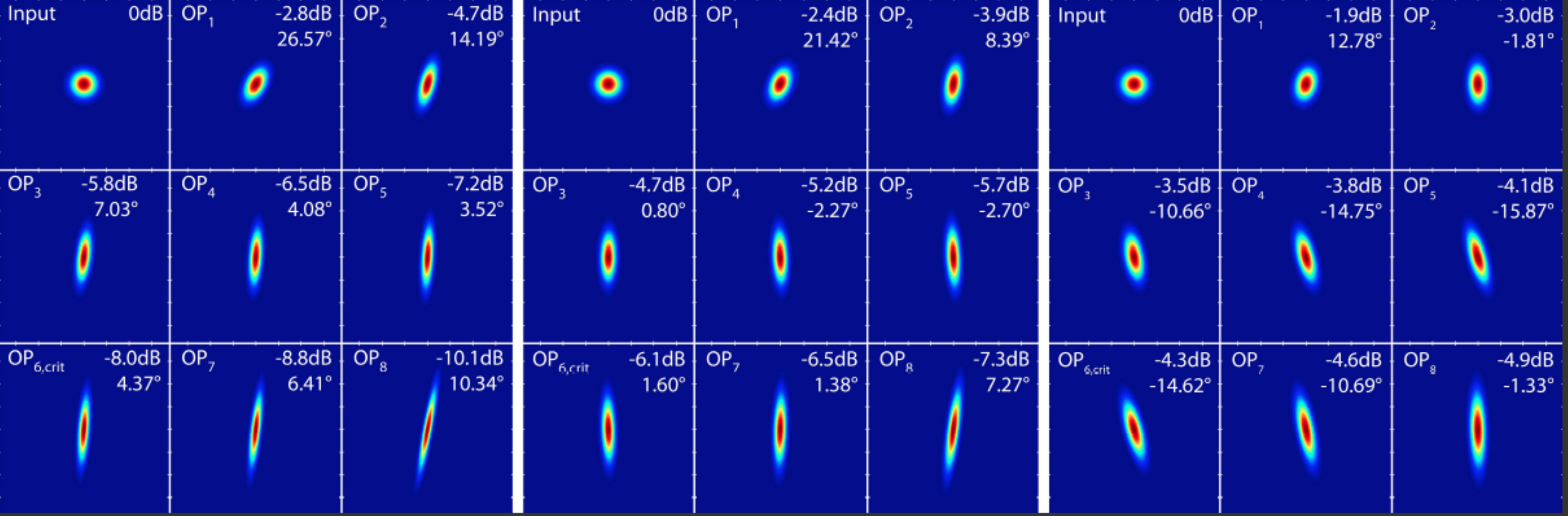}
\caption{ Noise transformations for {$\Omega=\gamma$}. \textbf{Left:}  ${\eta_{\rm esc} = 1}$ (loss-less). Due to the cavity dispersion at high frequencies no OP yields amplitude  quadrature squeezing.  \textbf{Middle:}  ${\eta_{\rm esc} = 0.9}$. In contrast to the loss-less case there still exist two OPs at which the squeezed quadrature matches the $X_1$-quadrature (one close to OP$_3$ and another close to OP$_{6,\rm crit}$).  \textbf{Right:} $\eta_{\rm esc} = 0.75$. Due to the high optical loss the squeezing is strongly degraded. At OPs close to OP$_2$ and OP$_8$ the achievable amplitude quadrature squeezing is about 3\,dB and 4.9\,dB, respectively.}
\label{fig:Wiggamma}
\end{center}
\end{figure*}

Fig.~\ref{fig:Wig01gamma} and Fig.~\ref{fig:Wiggamma} show the noise transformation of a shot noise limited driving field at two different Fourier frequencies,  $\Omega=0.1\gamma$ and $\Omega=\gamma$ respectively. For both cases we considered three values for the escape efficiency ($\eta_{\rm esc} = 1$, 0.9 and 0.75). The left 9 tiles of Fig.~\ref{fig:Wig01gamma} ($\Omega=0.1\gamma$) and Fig.~\ref{fig:Wiggamma} ($\Omega=\gamma$) show the Wigner functions obtained for $\eta_{\rm esc}=1$. One can see that the squeezed quadrature (e.g.~the semi-minor axis of the noise ellipse) rotates towards the \mbox{$X_1$-quadrature} when approaching the critical point OP$_{6,\rm crit}$. When the intra-cavity power is further increased (OP$_7$ and OP$_8$) it rotates in opposite direction back towards the \mbox{$X_2$-quadrature}. For $\Omega=0.1\gamma$ (Fig~\ref{fig:Wig01gamma}, left) being well inside the bandwidth of the KNLC  the squeezed quadrature matches the \mbox{$X_1$-quadrature} at the critical point (OP$_{6,\rm crit}$). Considering $\Omega=\gamma$ (Fig~\ref{fig:Wiggamma}, left) the squeezed quadrature only approximates  the \mbox{$X_1$-quadrature} at OP$_{6,\rm crit}$. This can be explained by the phase-space rotation caused by  a detuned cavity at higher frequencies.  

The Wigner functions obtained for  $\eta_{\rm esc}< 1$ (middle and right blocks of Fig.~\ref{fig:Wig01gamma} and Fig.~\ref{fig:Wiggamma}) show a qualitatively different behavior of the phase-space rotation. The first essential difference is that for $\Omega=0.1\gamma$ in all considered cases the squeezed quadrature does not match the \mbox{$X_1$-quadrature} if the  KNLC is operated at its critical point. Again, this fact explains the enhanced noise at low frequencies in the corresponding spectra (refer to Fig.~\ref{fig:specx1}). The second difference is that the squeezed quadrature oversteps the \mbox{$X_1$-quadrature} at a certain OP and starts to rotate back again at OPs corresponding to higher intra-cavity powers. That means the semi-minor axis of the noise ellipse coincides with the \mbox{$X_1$-quadrature} at \emph{two} OPs that both depart from the critical point. Both potentially yield a purely optical reduction of laser power noise or a bright amplitude squeezed state, respectively. Furthermore, from the comparison of the Wigner functions for $\Omega=0.1\gamma$ and $\Omega=\gamma$ one can deduce that the squeezing level, or more generally the noise reduction, can be optimized for a certain frequency by a proper choice of the OP.  Looking  at the left block of Fig.~\ref{fig:Wig01gamma}  the squeezed quadrature optimally approximates  the $X_1$-quadrature at a frequency of $\Omega=0.1\gamma$ if the KNLC is operated at OP$_{6,\rm crit}$. In contrast, for $\Omega=\gamma$ (left graph of Fig.~\ref{fig:Wig01gamma}) the best approach is obviously obtained at a OP being close to OP$_5$.

\begin{figure}
\begin{center}
\includegraphics[scale=0.3]{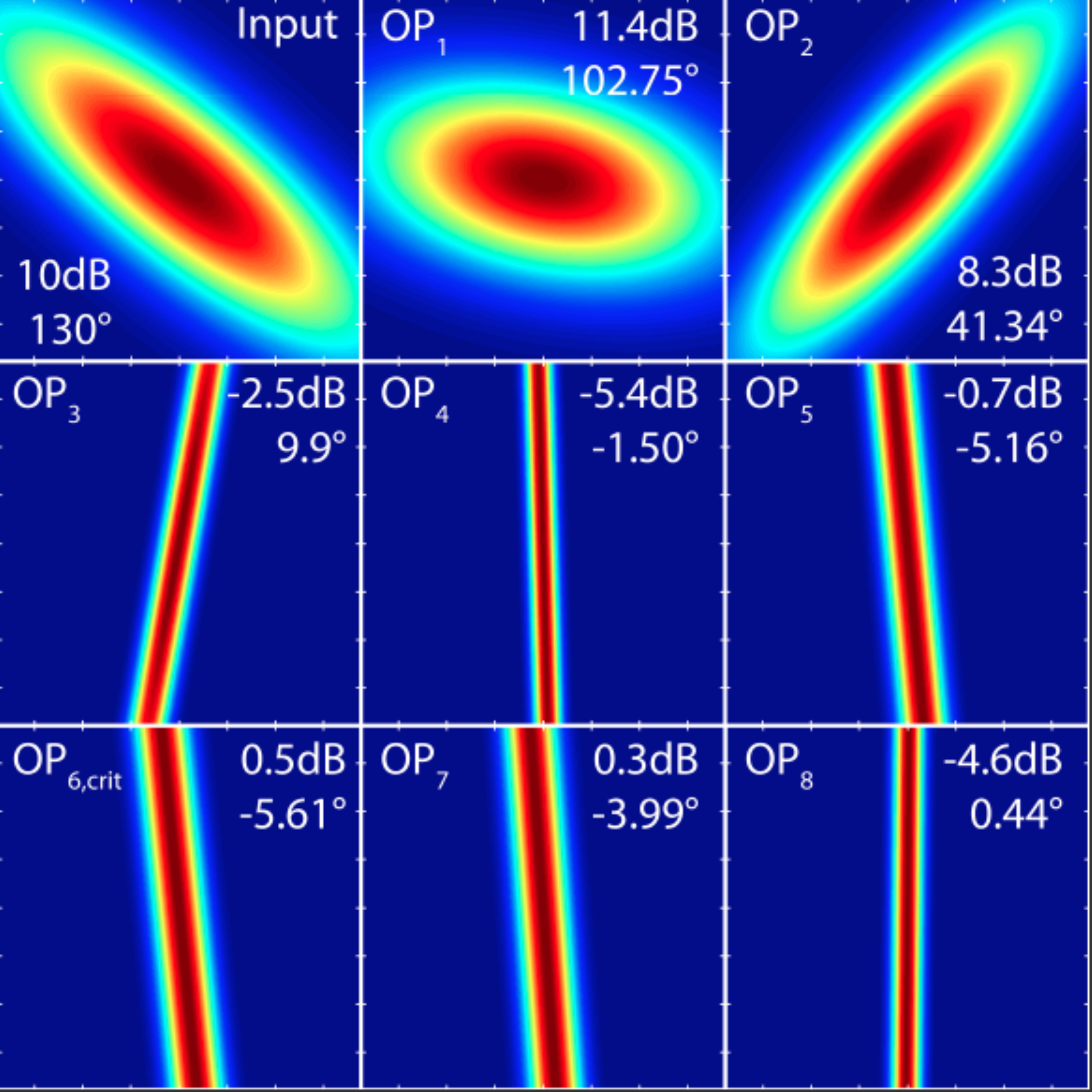}
\caption{ Transformations of a driving field (upper left graph) with an unbalanced noise distribution in phase-space. The semi-major axis of the input noise ellipse corresponds to a noise level of 20\,dB above the shot noise reference, the semi-minor axis to 10\,dB above shot noise. Again, for every Wigner function the angle and the vacuum normalized variance of the state's lowest noise quadrature are given. The KNLC has a escape efficiency of \mbox{$\eta = 0.90$}. The considered frequency is {\mbox{$\Omega=0.1\gamma$}.} }
\label{fig:arbWigeta7501}
\end{center}
\end{figure}

Fig.~\ref{fig:arbWigeta7501} illustrates  the noise transformation of a driving field showing significant classical noise, i.e.\ being in a highly mixed state. This field is constituted just as in the previous section (refer to Fig.~\ref{fig:arbspecx1}).  The KNLC has an escape efficiency of $\eta_{\rm esc}=0.9$. The  frequency considered is $\Omega=0.1\gamma$. Again, it can be deduced that there exist two OPs (one close to OP$_4$ and another close to OP$_8$) that yield a noise reduction in the $X_1$-quadrature of the reflected mean field. Although the driving field exhibits considerable classical noise, a reduction even below shot noise can be achieved. From the comparison with the middle plot of Fig.~\ref{fig:Wig01gamma} one can deduce that the orientation of the input field's noise ellipse has a significant influence on the squeezed quadrature of the transformed noise, but only at OPs corresponding to intra-cavity powers smaller than $0.5P_{\rm res}$. For higher intra-cavity powers (OP$_4$ to OP$_8$) where the (anti-)squeezing levels increases, the squeezed quadrature can be found to be almost the same as in the case of a shot noise limited driving field.

\section{Optimisation in the case of loss}

\begin{figure}
\begin{centering}
\includegraphics[scale=0.71]{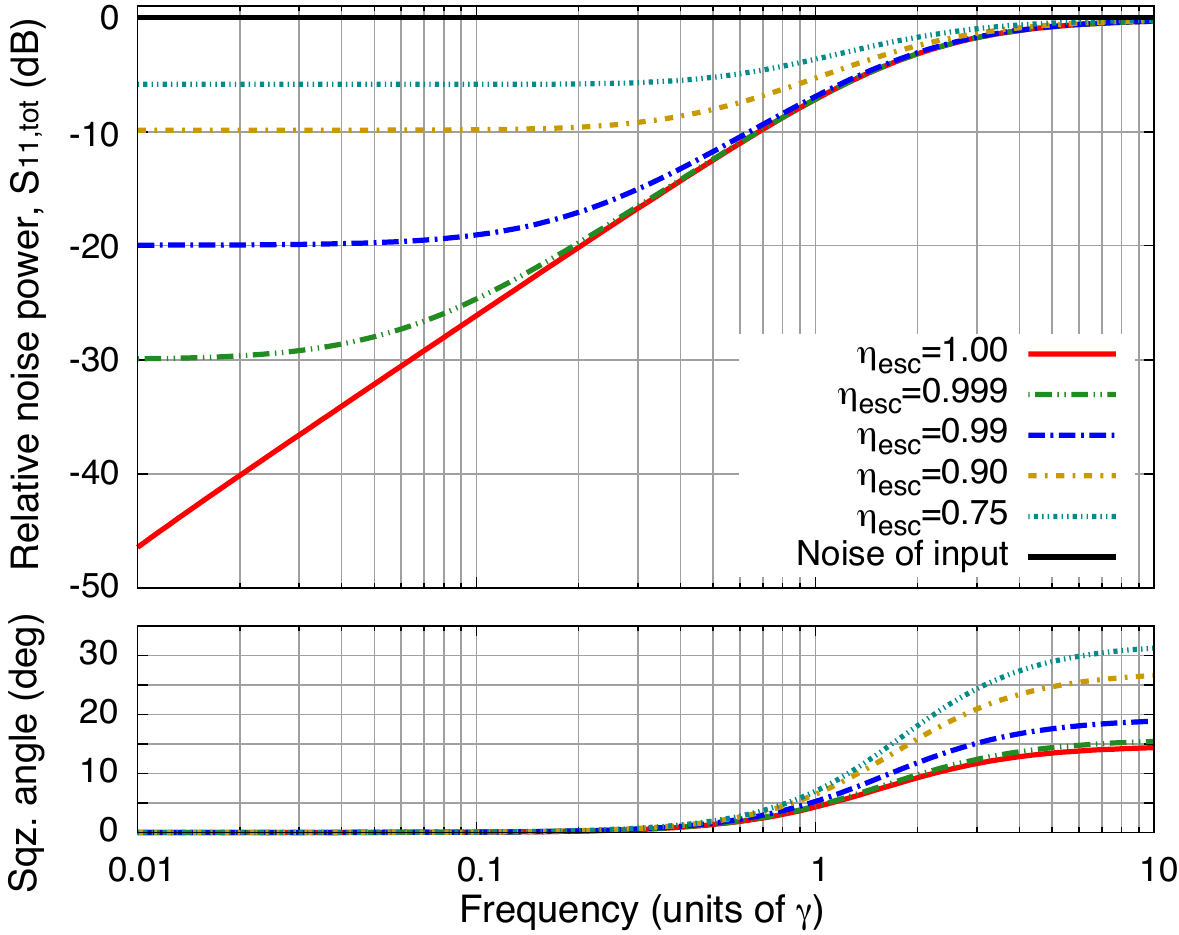}
\caption{ \textbf{Top:} Amplitude quadrature noise spectra for a {shot-noise limited input field reflected off the KNLC, which is operated at the optimum operation points, respectively, in order to provide maximum squeezing levels for different intra-cavity loss values.}  The squeezing levels approach the value ${10\log_{10}(1-\eta_{\rm esc})}$ at low frequencies. \textbf{Bottom:} Frequency dependence of the squeezing angle (quadrature angle of lowest noise). The OPs are chosen such that this angle is zero at low frequencies in all cases.}
\label{fig:specx1-opt}
\end{centering}
\end{figure}

From the phase-space representation of the noise transformation (Figs.~\ref{fig:Wig01gamma} and \ref{fig:Wiggamma}) we found that also a critical KNLC with internal loss can yield a noise reduction in the amplitude quadrature of the reflected mean field if the OP is chosen properly. By the choice of the OP the noise reduction can be optimized at a certain frequency.  Here we focus on low sideband frequencies being \emph{below} the cavity half-linewidth $\gamma$. This regime is of great importance since passive filtering with standard optical cavities can provide a noise reduction only at frequencies \emph{above} the cavity half-linewidth. Remember that there are two potential OPs yielding a noise reduction at low frequencies. We found that the OP that corresponds to higher intra-cavity powers yields slightly better noise reduction. 
The spectra obtained at the respective optimized OP for a shot noise limited driving field are shown in Fig.~\ref{fig:specx1-opt}.  The lower graph shows the frequency dependence of the quadrature angle yielding the lowest noise. As intended, at low frequencies it coincides with the amplitude quadrature of the reflected mean field. The achieved squeezing levels (top graph) are solely limited by the intra-cavity loss and approximate the value $10\log_{10}(1-\eta_{\rm esc})$ at low frequencies. Although the optimized OPs depart from the critical point the degradation of the squeezing level is solely caused by the vacuum noise contribution corresponding to the optical loss. 

\begin{figure}
\begin{centering}
\includegraphics[scale=0.71]{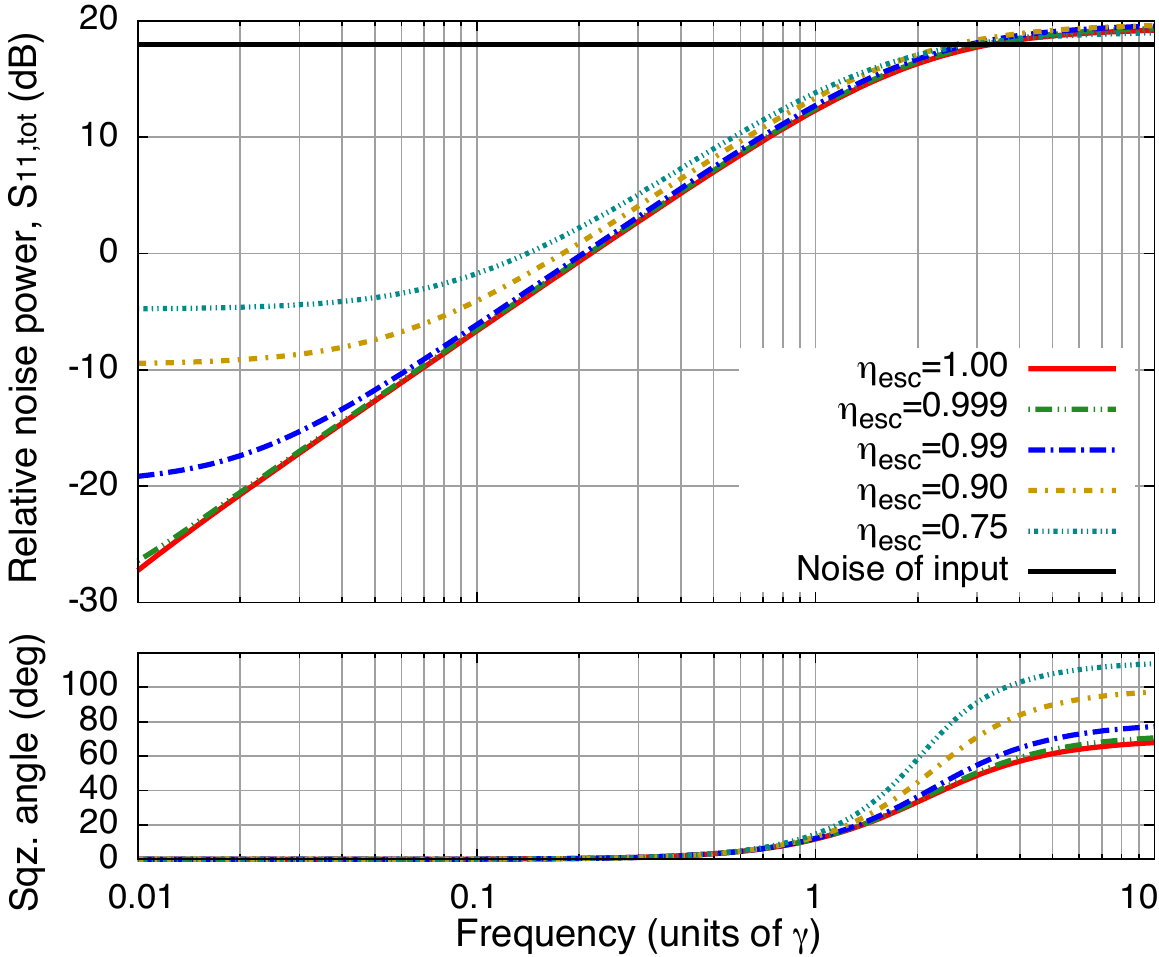}
\caption{ Amplitude quadrature noise spectra with optimized noise reduction as in Fig.~\ref{fig:specx1-opt}, but here for the input field previously considered in Figs.~\ref{fig:arbspecx1} and \ref{fig:arbWigeta7501}.}.
\label{fig:Arb-specx1-opt}
\end{centering}
\end{figure}

Similarly, we consider the optimization for the input field showing classical noise as already considered for Figs.~\ref{fig:arbspecx1} and \ref{fig:arbWigeta7501}. The resulting spectra are shown in Fig.~\ref{fig:Arb-specx1-opt}. Despite significant driving noise considerable squeezing levels can still be achieved at low frequencies.  In the presence of optical loss the squeezing level at low frequencies is still limited to a value of $10\log_{10}(1-\eta_{\rm esc})$, i.e.\ classical  noise of the driving field does not limit the squeezing but its {bandwidth}. In view of a  purely optical passive reduction of laser power noise in the classical regime, a considerable reduction at low frequencies is always possible, independent of the input fields noise distribution.

\section{Comparison with experimental data} \label{sec:exp}
\begin{figure}[t]
\begin{centering}
\includegraphics[scale=0.71]{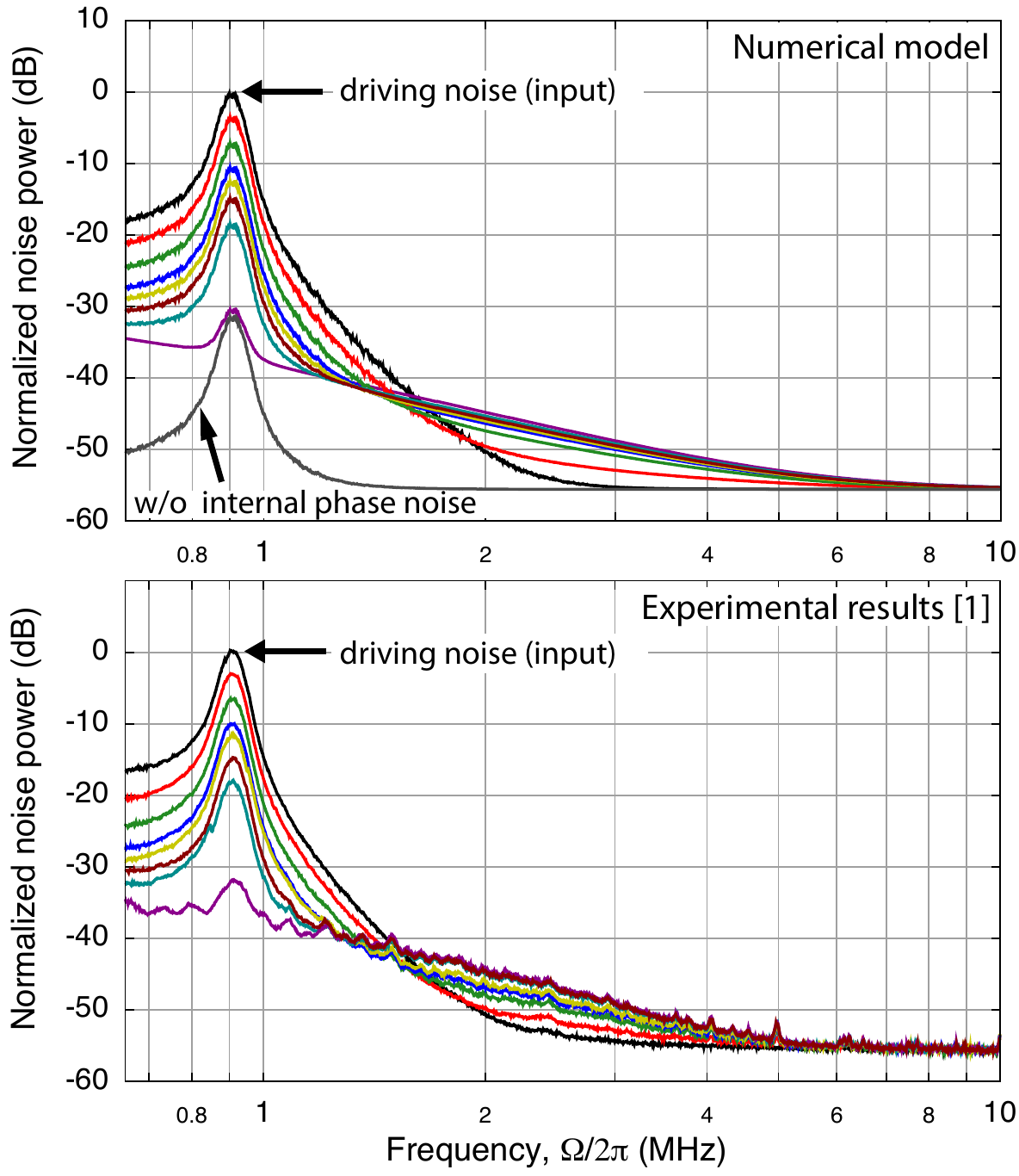}
\caption{ Comparison of our modeled  (top) with the measured noise reduction spectra (bottom) obtained in \cite{KhalaidovskiPRA80} {for different operation points of the KNLC, normalized to the peak input noise. Our model assumes an additional intra-cavity $1/f$-phase-noise in order to achieve the qualitative agreement shown.} The grey curve in the top graph is obtained instead of the magenta curve, if no additional phase noise is considered.}
\label{fig:exp-vs-sim}
\end{centering}
\end{figure}
Finally we model the KNLC that was used in \cite{KhalaidovskiPRA80} for a passive, purely optical reduction of laser power noise.  Its coupling mirror power reflectance was  $R_{\rm c}=0.983$. The intra-cavity round-trip loss was estimated to 0.5\,\% resulting in an escape efficiency of $\eta_{\rm esc}\approx0.77$. The KNLC's half-bandwidth was about $\gamma \approx 2\pi\cdot4.5\,{\rm MHz}$. The pump beam was guided through a modecleaner  ring cavity acting as spatial and low-pass filter (refer to Fig.~2 in \cite{KhalaidovskiPRA80}). The filtered beam had a power of about 750\,mW and was coupled to the KNLC. It was shown in \cite{KhalaidovskiPRA80} that for this input power the KNLC was very close to its critical state. The reflected beam was guided  through the mode cleaner a second time and eventually detected with a  photo diode realizing a measurement of the reflected mean field's amplitude quadrature noise. 

In order to model the experimental situation of  \cite{KhalaidovskiPRA80}, first, the characteristics of the input field need to be taken into account. Thus, we have performed a spectral analysis of the power noise of a free running Nd:YAG laser identical to that used in \cite{KhalaidovskiPRA80}. 
Additionally, we performed a tomographic noise analysis at the frequency of the laser's relaxation oscillation using a balanced homodyne detector. The tomography has revealed that the semi-major axis of the reconstructed noise ellipse deviates by roughly $\vartheta=10^\circ$ from the mean field's amplitude quadrature (i.e.  \mbox{$X_{\vartheta-10^\circ} = X_1$}). Furthermore, the noise level in the  \mbox{$X_{\vartheta+90^\circ}$-quadrature} corresponding to the semi-minor axis of the noise ellipse has been found to be about 33\,dB below the level in the \mbox{$X_{\vartheta}$-quadrature}. 
In our noise model we use the appropriately scaled power noise measurement as a description of the noise spectrum in the \mbox{$X_\vartheta$-quadrature}. 
Also in accordance with measurement results we approximated the noise in the \mbox{$X_{\vartheta+90^\circ}$-quadrature} by a cavity pole function having a $1/f^2$-scaling above the laser relaxation oscillation. Furthermore, the orientation of the noise ellipse in phase-space is assumed to be $\vartheta=10^\circ$ for all frequencies. In order to account for the modecleaner in the experimental setup, the input noise data are multiplied by a another pole function describing the modecleaner's low-pass behavior before the transformation introduced by the KNLC is calculated.

Second, we have verified that the assumption of a linear noise transformation is still appropriate for the power and frequency regime investigated in  \cite{KhalaidovskiPRA80}. Although the noise level in the \mbox{$X_{\vartheta}$-quadrature} of the driving field is at the laser relaxation oscillation frequency several orders above shot noise, our time-domain simulation does not show any harmonic frequencies. 
Instead, we find that for the modeled critical KNLC ($P_{\rm in} = 750\,\rm mW$, $\eta_{\rm esc} = 0.77$, $\gamma=2\pi\cdot4.5\,{\rm MHz}$) the non-linear property of the transfer function in the frequency regime above $\Omega=0.1\gamma$ only becomes significant for input noise levels 5 orders of magnitude above the experimentally found peak value. 
Finally, the field reflected off the critical KNLC is multiplied by the pole function and attenuated to about 150\,mW in order to model the modecleaner low-pass filtering and the photo-electric detection scheme used in  \cite{KhalaidovskiPRA80}, respectively.  

Fig.~\ref{fig:exp-vs-sim} compares the modeled $X_1$-noise-reduction spectra (top) with the experimental results (bottom). As in \cite{KhalaidovskiPRA80}, seven exemplary OPs are considered. They are chosen with respect to the noise reduction at the laser's relaxation oscillation frequency. As already stated in \cite{KhalaidovskiPRA80}, the experimental noise reduction spectra show additional noise in the mid-frequency  range. 
{We have verified that this noise is not due to the Kerr effect since it also appears when the Kerr effect is switched off by rotating the light's polarization by 90$^\circ$.}  
Presumably, this excess noise is thermally driven (e.g.~by thermo-refractive noise~\cite{BraginskyPL2000})  as the intra-cavity powers reached values in the order of 100\,W at a waist size of about $30\mu\rm m$. In our model we assume an intra-cavity $1/f$-phase-noise that scales with the square of the intra-cavity power. The strength of this phase noise is fitted such that the magenta curve is obtained from the spectrum yielding the best noise reduction (grey curve in Fig.~\ref{fig:exp-vs-sim}). This phase noise model is used for all other curves.  Doing so it is possible to model noise reduction spectra being in an excellent agreement with the experimental results. \\

\section{Summary}
We have theoretically investigated the noise transformation of a critical KNLC based on a rigorous treatment in a time-domain simulation. We have eventually aimed at the modeling of the passive laser noise reduction with such a cavity, as observed in \cite{KhalaidovskiPRA80}. The comparison of our modeled laser power noise reduction spectra showed an excellent agreement with the experimental results of the work mentioned when adding a $1/f$~excess phase noise. 
{
Our model revealed that the knowledge of both, the optical loss inside the KNLC as well as the phase space distribution of the input laser noise are crucial for a correct description. In our time-domain simulation of the experimental data of \cite{KhalaidovskiPRA80} we have not observe the transformation of input noise to other frequencies, i.e.  linearized equations would have been equally valid. We therefore restricted our entire theoretical analysis of the influence of optical loss and input noise phase space distribution presented here to the regime of linear transfer functions.
Our analysis has shown that a KNLC is generally able to provide a noise reduction beyond shot noise, even in the presence of optical intra-cavity loss and classical driving noise. We have found that the noise reduction in the amplitude quadrature of the reflected mean field can be optimized by the choice of the KNLC operating point.  The {noise reduction} level at frequencies much smaller than the KNLC's half-bandwidth $\gamma$ has not been found to be limited by the noise of the driving field, but solely by the intra-cavity loss. 
The bandwidth in which significant squeezing levels can be achieved, however, has turned out to be limited by the amount of classical noise carried by the driving field. We believe that the presented investigations and our numerical model have a high potential to estimate parameters required in existing and future experiments aiming at a purely optical reduction of laser noise or at the squeezing of quantum-noise based on a KNLC. A future experiment with considerably stronger driving noise than present in \cite{KhalaidovskiPRA80} may show whether our approach is also able to correctly model nonlinear transfer functions and the generation of non-Gaussian noise through a critical KNLC.
}

\section*{Acknowledgements}
This work was supported by QUEST, the center of excellence of the Leibniz Universit\"at Hannover, and the European Commission's Seventh Framework Programme (FP7/2007-2013) under the Grant Agreement 211743 (Einstein Telescope Design Study). Henning Rehbein, Nico Lastzka and  Jan Harms are acknowledged for fruitful discussions.\\

\end{document}